# Hierarchy of knowledge translation: from health problems to ad-hoc drug design

David Fajardo

Doctorado en Ciencias Medicas y de la Salud

Universidad Nacional Autonoma de Mexico

Victor M. Castano

Centro de Fisica Aplicada y Tecnologia Avanzada

Universidad Nacional Autonoma de Mexico

Abstract

An innovative approach to analyze the complexity of translating novel molecular entities and nanomaterials into pharmaceutical alternatives (i.e., knowledge translation, KT) is discussed. First, some key concepts on the organization and translation of the biomedical knowledge (paradigms, homophily, power law distributions, hierarchy, modularity, and research fronts) are reviewed. Then, we propose a model for the knowledge translation (KT) in Drug Discovery that considers the complexity of interdisciplinary communication. Specifically, we address two highly relevant aspects: 1) A successful KT requires the emergence of organized bodies of inter-and transdisciplinary research, and 2) The hierarchical and modular topological organization of these bodies of knowledge. We focused on a set of previously-published studies on KT which rely on a combination of network analysis and computer-assisted analysis of the contents of scientific literature and patents. The selected studies provide a duo of complementary perspectives: the demand of knowledge (cervical cancer and Ebola hemorrhagic fever) and the supply of knowledge (liposomes and nanoparticles to treat cancer and the paradigmatic Doxil, the first nano-drug to be approved).

Keywords: Biomedical knowledge, translational hubs, trading zones, scientific cultures.

# 1. Introduction

"From bench to bedside" is the motto of Translational Science (TS).[1] The fundamental idea behind TS is ultimately to translate the basic science discoveries into a better health status for the population.[2] Particularly, TS in Drug Discovery consists in transforming the knowledge on biomolecular entities -like small molecules, nucleic acids or proteins- and their interactions into drug therapies.[3] This translation of biomolecular knowledge into therapies is becoming an increasingly difficult and complex task. A productivity crisis in pharmaceutical research and development (R&D) has been extensively reported.[4] This crisis consists, among other things, of a decline in the number of New Molecular Entities approved, and an increase of the attrition of R&D projects, development times and R&D expenditures.[4] According to Pammolli, Magazzini and Riccaboni, the explanation of this innovation crisis is that the R&D efforts are concentrated in areas with a potential of very high sales per year but with the lowest probability of success. [4] These areas with a low probability of success are aimed to develop new drugs through unexploited biomolecular mechanism to treat complex diseases like cancer and obesity.[4] These innovation trends could be boosting the increase in size and complexity of biomedical literature, making it difficult to translate the basic knowledge in innovations.

The difficulty of translating the biomedical knowledge into treatments also could be observed in the unbalance between the huge quantity of published papers and the shortage of new approved drugs.[5] Moreover, the increasing complexity of the biomedical information could be tackling the construction of meaningful ideas and their transformation into innovations. That is, knowledge translation in Drug Discovery is becoming a big data problem.[6, 7] In previous investigations, we have shown how the combination of biomedical-bioinformatics and scientometrics could be useful to handle the complexity of biomedical knowledge and its translation into innovations.[8-11] These two fields have different approaches on the information. On the one hand, biomedical bioinformatics processes, organizes, structures, integrates and represents data in order to obtain meaningful and useful information that could be used, for example, to discover new drugs, make medical decisions or improve health care quality.[12] On the other hand, scientometrics focuses on quantifying the communication of scientific information.[13] Differently, our approach focuses on the communication of biomedical information among researchers and simultaneously organizes, structures and represents the communicated information using bioinformatics tools.[8-11] Particularly, knowledge translation is understood as a communication process between the basic and the clinical research levels.[14] In this process, a new interdisciplinary or translational language emerges.[14] We propose that the translational language and the intellectual community that produce it are both fundamentally necessary for the transformation of the huge mass of data obtained from the experimental basic research into the longed pharmaceutical treatments. Therefore, this new approach pays special attention to the organization, structure and representation of the knowledge. The bioinformatics component of our research strategy allows us to analyze the emergence of this new interdisciplinary language.

In this paper, we extensively discuss this new approach. In the second section, key concepts like scientific cultures, some network properties and interdisciplinary communication are briefly reviewed. In the section 3, a model for KT in Drug Discovery as a complex process of interdisciplinary communication is presented. In section 4, we will discuss four previously studied instances of knowledge translation processes. Two of these studies analyze the knowledge translation process of the research on specific health problems: cervical cancer[8] and Ebola[11]. The other two studies analyze this process since the perspective of the R&D of medical technologies. One study compares the NT process in two types of nanotechnologies to treat cancer: liposomes versus metal nanoparticles.[9] The other study focuses in the development of Doxil,[10] the first approved nanodrug.[15]

## 2. A brief review of some key concepts to the study of KT in Drug Discovery

### 2.1 Paradigms as scientific cultures

As defined by Thomas Kuhn, a **paradigm** is a scientific achievement that have two characteristics: 1) It is so relevant that a community of adherents is organized around it and 2) It leave an important quantity of unsolved problems to be worked by this community.[16] Importantly, the paradigms set a system of standards for scientific practices to be followed by the members of a particular scientific community.[16] Moreover, the paradigmatic standards determine what is to be observed, how the observations should be performed, how questions should be asked and how the results should be interpreted.[16] The kuhnian paradigm is a polysemic concept. It includes simultaneously the achievement, the standards and the community.[17]

Paradigms are particular types of intellectual cultures[17] and, importantly, they are associated with specialized language.[18] These specific languages determine the way the information is communicated, presented and interpreted inside a community of researchers. [18] Moreover, knowledge translation can be understood as a communication process where a new interdisciplinary language is built.[14] That is, scientific languages represent a coherent system of meanings, commitments and grammar rules that defines the membership in the communities. When two or more scientific cultures interact with each other, a sub-system of communication must be established, i.e., a trading zone.[17] These sub-systems are territories that belong to neither of the interacting communities but allow the exchange of concepts, methodologies, tools and evidence.[17] So, trading zones would allow the interdisciplinary communication without the need to altering the fundamentals of each culture.

Paradigms are important for the study of KT in drug discovery because they determine the way that NT actually takes place. For example, the discovery of the human papillomavirus (HPV) as the necessary cause of cervical cancer led the NT process toward the development of vaccines and HPV tests[8]. This example is discussed more extensively in section 4.

### 2.2. Key concepts: Homophily, power laws, hierarchy, modularity and research fronts in literature networks

Scientific journals are still the most important media of communication for the scientific community. The papers published in these journals build a complex citation network that shows both the organization and evolution of the scientific community and the knowledge itself.[19] That is why the analysis of literature networks is fundamental to the understanding of the knowledge translation process. Similarly, networks of patents provide relevant information about the development of technologies.[20] In this sub-section some relevant properties of these kinds of complex literature networks are briefly reviewed. The review of these properties is aimed to introduce the conceptual basis of the research fronts which are understood as the convergence of clusters of papers, specialized community of researchers and jargon. Research fronts are particularly important to the analysis of NT as a communication process, as it is discussed below.

**Homophily** is the trend of certain nodes (the objects represented in a network: persons, proteins, papers, etc) to preferentially establish connections with others nodes of similar features.[21] People tend to build relationship almost exclusively with people that are economically, socially and culturally akin.[21] Scientific literature and scientist coauthor networks are not different. Scientists preferentially collaborate with colleagues that belong to the same scientific culture.[22] Papers normally cite other manuscripts with very similar content. Importantly, according to McPherson, Smith-Lovin and Cook "The pervasive fact of homophily means that cultural, behavioral, genetic, or material information that flows through networks will tend to be localized. Homophily implies that distance in terms of social characteristics translates into network distance".[21] Therefore, homophily is the fundamental property that allows us to map the KT process. That is, basic and clinical research tends to be located in a well defined regions in the citation networks.[8-11, 23] Translational research would occupy an in-between topological position in the network connecting the basic with the clinical research.[8-11, 23] Similarly, papers that are in the same observation scale (molecular, cellular, organism, population) tend to stick together.[11] Moreover, it is completely feasible to identify research communities that work on common problems with similar approaches using network analysis tools like those based in the Newman's modularity.[8-11, 20]

The distribution of citations and information among the papers and patents follows a **power law function** (were a variable is the power of other).[24] In the case of citations, a power law distribution implies that a very few number of papers (or patents) receive a huge quantity of citation and the most of manuscripts are few or never cited. In the case of information, the power law distribution means that there are few ideas, methodologies or scientific achievements that are constantly mentioned in the scientific literature and a huge diversity of new ideas, methodologies or results that are poorly mentioned in the literature.[24] The power law distribution of citations and information has important methodological implications to the analysis of the KT as a communication process. The analysis of top cited (or with the highest degree) papers and their citation network can provide the most relevant information about how a research area is organized and evolves and how the research levels are inter-communicated.[8-11, 19 20, 23, 25-27] Some authors arbitrarily choose the 45 or 100 most frequently cited papers.[25, 26] While others authors perform the main path analysis to identify the putative path with the broadest flow of information.[27] In our previous research, we have chosen the 20% most frequently cited papers. This quantity of top cited papers guarantees to get the most of the communication process through the citation networks and the most of the communicated information.

Moreover, the power law distribution of citations and information in the scientific literature (including patents) are strongly related to the idea of paradigms. Generally, the most cited papers are related to 1) methodologies and research tools that establish the standards of scientific practices of the research areas[28] and 2) the scientific achievement that comprise the paradigmatic explanation of a problem.[19] Therefore, the network of top cited papers in a particular research area is strongly related to its paradigmatic research structure.

**Modularity** is another important property of complex networks that refers to the strength of the division of a network in modules or clusters.[29] Clusters (modules) are groups of nodes highly connected each other but poorly connected to those nodes that belong to other groups.[29] Interestingly, modularity and homophily are related: the nodes in a cluster share some properties and information that are not common to the nodes belonging to other clusters. That is, clusters are functional, meaningful and identifiable parts of the network. For example, metabolic networks are divided in modules related to different pathways like carbohydrate metabolism or amino-acid metabolism.[30] Moreover, modules of papers (**research fronts**) are related to communities of researchers that share in some measure common research interests, objects of study, jargon and standards of scientific practices.[31-32] That is, research fronts could be related to certain scientific cultures or sub-cultures.

Importantly, a culture focused in a very specific object of study could be embedded in a particular paradigm (a broader culture) and a paradigm in turn could be part of a research style (a very general way of practicing science), forming a hierarchy. In this regard, a useful concept is the **network hierarchy** proposed by Albert, Ravasz, and Barabási to explain the coexistence of modularity and the power law distribution of links in complex networks.[33] This model consists of assembling of network modules in larger modules that are in turn forming bigger modules and so on.[33] However, the way that modules are assembled is restricted according a scaling law wherein the clustering coefficient of a node is inversely proportional to its degree (number of links).[33] That is, modules are inter-communicated by nodes with a very high degree and a small clustering coefficient.[33] These classes of nodes are named hubs. By combining hierarchy and homophily a model of interdisciplinary communication can be glimpsed. In this model, the scientific cultures embodied in research fronts or clusters of research fronts are connected by interdisciplinary hubs. We above mentioned that homophily means that the knowledge that flows along the literature networks is located. Therefore, the knowledge that pass through the hubs does not belonged to any of the cultures. Instead, a negotiation process (trading zones) between the cultures would take pace in these hubs. This process is discussed in the following section.

**2.3 The KT inside and outside the biomedical field**

Our approach considers three levels of KT. The first level communicates the biomedical knowledge with the clinical knowledge. These are two completely distinct scientific cultures. The biomedical knowledge has a strong focus on the diseases and its argumentation is accumulative (like building blocks) while clinical knowledge is centered on the patient and its argumentation is competitive (For example, to establish which would be the most suitable health intervention for a particular case).[34] The second level of KT communicates the constituent disciplines of the biomedical knowledge. Five major disciplines have been proposed to make up the biomedical field: Epidemiology, clinical trials, diagnosis (includes screening), therapy development and pathogenesis.[34] This proposal has been partially supported by an analysis of the co-citation network of the 100 top most cited papers on cancer. This analysis showed that the leading cancer research is organized in six clusters: namely, epidemiology, clinical trials, diagnosis, molecular etiology, microarrays and targeted therapies.[23] Our previous studies on the structure of cervical cancer and Ebola research show a similar disciplinary division as we describe below.

A third level of KT occurs in the development of specific treatments. At this scale of observation the KT is the connection among the stages of invention (wherein the scientific and technological basis of a product are established), innovation (i.e., the development of a new treatment) and imitation (the development of products strongly based in the original).

The KT in Drug Discovery (the transformation of chemical substances into drug therapies) particularly under the paradigm of the rational design, is a process that fundamentally takes place inside the biomedical field. That is, in order to get a newly approved drug in the market, the knowledge on the substances must go through four of these major disciplines of the biomedical science, from basic knowledge of the causes and mechanisms of diseases (etiology and pathogenesis), to the development of drug therapies or tools for diagnosis or screening. Then, the Investigacional New Drug should be evaluated in the different phases of clinical trials in order to get the approval. Finally, it is necessary to evaluate the impact of the new treatment (also, diagnosis tools) in the population level (epidemiology).

**3. A model of KT as a communication process among research fronts**

Two are the central elements of the proposed model of KT: Research fronts and translational hubs. Research fronts are identifiable by clustering the networks of scientific literature.[31, 32] They are related to scientific cultures like biomedical disciplines ( epidemiology, therapy development, pathogenesis, etc.), therapeutic approaches (gene therapy, induced hyperthermia, drug therapy) or communities working on the development of a specific product.[8, 9, 11, 19, 23] Moreover, clustering can identify the stages of invention, innovation and imitation.[10] Importantly, these research fronts are hierarchically organized similarly to the Russian dolls (Figure 1). That is, at the highest level there are the clinical knowledge and the biomedical knowledge. Inside of the biomedical knowledge there are located the fundamental biomedical disciplines like molecular etiology, diagnosis, therapy development or diagnosis. In turn, the molecular etiology can be organized in small communities aimed to the study of specific molecular mechanism. Also, inside of the drug development discipline could be found different therapeutic approaches.  Finally, within a specific therapeutic approach or inside of the development of a product line it is possible to identify communities of inventors, innovators and imitators.

Translational hubs are regions in the literature network that connect the research fronts at different scales (Figure 1). These regions are highly hierarchical and because of the homophily they contain a sort of a mix of the information contained by the communicated fronts. However, the dynamics inside the translational hub are much more complex. Translational hubs are the topological location of the trading zones, which implies the emergence of a transdisciplinary language and a negotiation process.  When two or more research fronts (scientific cultures) interact each other, they interchange terms, data, arguments, tools, etc. The interchanged objects are frequently used and understood differently by the giver and the receiver research fronts according to their respective discursive structures.  This interchange of atomized elements -and commonly alienated of their original meanings- is what define the multidisciplinary collaboration among research fronts. On the other hand, the emergence of any language requires an hierarchical articulation of the communicated elements. [35] That is, the interchanged elements are combined to form more sophisticated concepts, and then arguments or conceptual relationships, and so on until a new interdisciplinary narrative and language is built. Moreover, the construction of a new interdisciplinary language implies a negotiation process among the scientific cultures to establish what elements are to be shared and how the conceptual relationship, the argumentation and the methodologies should be articulated. [17] The importance of the negotiation process for the KT is that the former partially deters what molecular mechanism or pathological process will inform the development of therapies; what therapeutic approach is going to be tested in the clinical trials, or what screening tools will be developed.

In the next sections the two key elements of the proposed model of KT (research fronts and translational hubs) are illustrated through the revision of five studied cases.[8-11] This instances represent different levels in the knowledge hierarchy: 1) cervical cancer research and Ebola research as examples at the biomedical knowledge level); 2) liposomes and metal nanoparticles to treat cancer as examples of therapy development), and 3) liposomal doxorubicin as an example of a sub-system of products.

There are some variations in the methodology used to study the previously mentioned instances, particularly between the study on the KT process in cervical cancer research and the others. However, this approach is consistent in the next four essential aspects: 1) A citation network model of the top cited papers is built, 2) in order to map the distribution of basic, translational and clinical papers in the model the proportion of clinical terms versus basic terms is calculated for every paper, 3) a clustering analysis is performed in order to identify the research fronts, and 4) The content of each identified cluster is analyzed.

The methodology produces a map of the research on a specific health problem, discipline or the development of a certain drug. This map identifies the main research fronts (scientific cultures or sub-cultures) and the structural relationships among them. Importantly, this model identifies where and how the KT process takes place.

**4. Knowledge translation since the perspective of the biomedical knowledge of health problems**

**4.1 The context of Ebola and cervical cancer**

Cervical cancer and Ebola represent two opposite classes of diseases: the chronic diseases (that persist over time) and the acute diseases (rapid onset, a short course, or both) respectively. This is important for the study of knowledge translation because the treatment of chronic disease over the time would require a deeper and more complex knowledge on the patient in comparison to the acute diseases. However, cervical cancer and Ebola share some features as health problems. First, both diseases mainly affect unprotected population in developing countries.[35, 36] Secondly, the affected persons are racially stigmatized in the case of Ebola fever[37, 38] and a there is a sexist stigma in the case of cervical cancer patients who commonly are irrationally accused of sexual misconduct.[39] The stigmas and the unprotected condition of the patients could be negatively impacting the development of a diversity of strategies to prevent or treat these diseases.[8, 11]

**4.2 The structure of KT in cervical cancer research and Ebola research**

The structure of the knowledge on these two health problems is simpler than the general proposed model depicted in the figure 1 (Compare fig. 1 with fig. 2). That is, not all the biomedical disciplines are enough developed to be identified as research fronts.[8, 11] Probably, it is due to their context, which is similar to the situation of the neglected diseases.

A clear division between the biomedical (disease centered) knowledge and the clinical (patient centered) knowledge is observed in cervical cancer research.[8] Furthermore, there is a very poor communication (inter-citation) between these two fundamental research styles.[8] Moreover, the biomedical knowledge on cervical cancer comprises mainly three disciplines, namely, molecular etiology, epidemiology and screening.[8]

The translational hub is fundamental for the structure of cervical cancer research. This hub communicates the molecular ethology of the disease with the epidemiology and the development of screening tolls (Fig 2).[8] But more importantly, this hub is intrinsically related to the research community that discovered that some variants of the human papillomavirus (HPV) are the necessary cause of the cervical cancer.[8]

On the other hand, there is not an identifiable body of clinical knowledge on Ebola.[11] Probably, this is due to the extreme acute and lethal nature of the disease not providing chance to produce knowledge on the ill patient.[11] Interestingly, the structure of biomedical Ebola research is highly reductionist.[11] That is, the molecular etiology of the disease is represented by three research fronts specialized in three respective viral proteins: VP35, VP40 and the glycoprotein.[11] As a matter of fact, the research on the viral glycoprotein is the biggest front and it strongly influence the development of therapies.[11] The reductionism in Ebola research could be partially explained by a combination of its social context with the fact that Ebola research is strongly militarized and affected by a national security approach in the United States.[11] The general structure of Ebola research is schematically represented in Figure 2. Besides the fronts focused in viral proteins there are fronts related to the disciplines of pathogenesis, epidemiology and therapy development.[11] The translational hub in Ebola research is located in the interaction among the glycoprotein research front, pathogenesis and drug development.[11] This hub is fundamentally aimed to the development of immunotherapies targeting the viral glycoprotein.[11]

**5. Knowledge translation since the perspective of the therapy development**

**5.1 Liposomes and metal nanostructures to treat cancer**

Liposomes and metal nanostructures are relevant examples of nanoscale technologies aimed to treat cancer. Liposomes are nanoscopic bubbles of lipid bilayer.[40] While, metal nanostructures (MN) are a plethora of different metal (mainly gold) nanoscale particles like nanorods, nanosell or nanocage.[41]

Liposomes are broadly versatile drug delivery systems that help to avoid the harmful side effects of drugs and increase the accumulation in the target tissue.[40] They can be combined with small drugs, antibodies or nucleic acids.[40] On the other hand, the extraordinary optical properties of MN in terms of light absorption, light scattering and fluorescence allow their use in imaging and photothermal therapy (the destruction of cancerous tissues by heating with light) to treat cancer.[41]

Importantly, these two types of cancer nanotechnologies are related to different explanatory models of cancer. Liposomes are comparatively more akin to the dominant paradigm of cancer as a diseases caused by an accumulative malfunctioning of the molecular machinery of the cells. That is, liposomes tend to be integrated in therapeutic strategies intended to impact or take advantage of the biomolecular dynamic of the cancerous cells. On the other hand, MN are relatively more independent of the biomolecular paradigm of cancer and more close to an explanation of cancer as a disruption in the structure of the tissues. That is, what is important in the use of MN is to reach the cancerous tissues and to physically destroy them. The affinity to different explanatory models of cancer could affect the KT process in these two cancer nanotechnologies being MN a much more radical technology than liposomes.

**5.2 The structure of KT in these two types of cancer nanotechnologies**

The research on liposomes and cancer is organized in six meaningful research fronts.[9] Two of the research fronts are focused on the combination of liposomes with doxorubicin (a small molecule), one of these front is aimed to the development of the drug while the other front consists of clinical trials testing the liposomal doxorubicin.[9] The other research fronts are related to different therapeutic approaches: gene therapy, induced hyperthemia, small interfering RNA, and drug therapy.[9] Moreover, the KT in liposomes and cancer research only occurs for liposomal doxorubicin while the other therapeutic approaches are still basic research.[9] This is strongly related to the development of Doxil (pegylated liposomal doxorubicin) which was the first nanodrug approved by the Food and Drug Administration of the United States.[9]

On the other hand, MN and cancer research do not display a KT process, i.e., it is sill basic research.[9] Also, the research on this cancer nanotechnology is not clustered, i.e., it consists of a single research front.[9] The differences in terms of structure and knowledge translation between liposomes and MN applied to cancer could be explained by two factors. First, MN represent a more radical invention in comparison with liposomes with would be relatively a more incremental or conservative technology.[9] Second, liposomes are a complementary technology that can be combined with different class of objects while MN are themselves the therapeutic element.[40]

**6 Knowledge translation at the scale of a sub-system of products: the liposomal doxorubicin formulations**

The liposomal doxorobucin formulations (LDF) represent an example of a sub-system of products. That is a set of products that follow a dominant design, i.e., a set of core features that are established as de facto standards.[42] A design becomes dominant when its features are so appealing that attract an important segment of the market and force the competitors to imitate the design.[42] In this case Doxil (PEGylated liposomal doxorubicin) is the dominant design for this type of formulations.[10] Importantly, LDF is also a subset of liposomes and cancer research. Therefore, the map of LDF research is the zoom in view of the two research fronts related to doxorubicin in the map of liposomes and cancer research above discussed.

The network model showed that LDF research is organized in fronts related to communities of inventors, innovators and imitators.[10] There was identified a research front of basic research wherein the technological and scientific basis of this type of formulations were established (the invention stage).[10] A second front is related to the research community that developed Doxil (the innovation stage) and it is connected to a third research front of clinical trials.[10] This clinical trials were aimed to firstly obtain the FDA approval to treat the Kaposi sarcoma and secondly to extend the use of Doxil to treat other cancers.[10] Following these fronts there are others that are related to 1) the combination of Doxil with others therapies[10] and 2) the improvement of liposomal doxorubucin formulation by mean of the use of thermosensitive liposomes and antibodies.[10] This last follower fronts constitute the imitation stage.

The front related to the development of Doxil represents the translational hub that connects the basic research (invention stage) with the clinical trials.[10] Interestingly, the innovators constitute an international community of collaborators aimed to integrate three different technological solutions: 1) nano-sized liposomes, 2) Use of pH or ammonium ion gradients for loading of the drug in the nanoliposomes, ans 3) the stealthness technology in order to avoid the reticuloendothelial system.[10, 15]

## 7. Conclusions

In this manuscript we propose an alternative to the linear model (the pipeline model, see [1]) of KT. Our model considers that KT is a hierarchical multiscale phenomenon that can be traced through a combination of network analysis and content analysis (text mining plus the study of the distribution of Medical Heading Subjects and Gene Ontology terms). Clustering analysis can identify research front at different scales ranging from research styles (biomedical knowledge and clinical knowledge) to molecular mechanism or a systems of products like the liposomal doxorubicin formulations. Moreover, KT can be located at specific region of the network model, namely translational hubs. We provided a few examples of KT at different observation scales. However, further research is required. Particularly, we consider that analysis the structure of KT in big health problems like cancers, obesity, or depression at different scales (health problems, biomedical disciplines, therapeutic strategies and so on) can provide fundamental information to a cleared understanding of KT as a complex communication process.

**Conflict of interest**
The authors confirm that this article content has no conflict of interest.

**Acknowledgment**
David Fajardo-Ortiz is partially supported by the CONACYT PhD scholarship.

**Figure 1.** Hierarchical model of the knowledge translation. Each rounded rectangle represent a research front. They are hierarchically organized in four levels (L1 to L4). The Clinical knowledge and the biomedical

knowledge are at the highest level of the organization of knowledge (L1). Inside the biomedical knowledge (L2) there are six of the main biomedical disciplines: Clinical trials, epidemiology, molecular etiology, pathogenesis, screening-diagnosis and therapy development. At the level 3 there are research fronts that could be specialized in certain molecular mechanisms or in specific therapeutic approaches. Inside a therapeutic approach (L4) it is possible to identify communities of inventors, innovators and imitators. The ovals with the TH acronym represent the translational hubs.

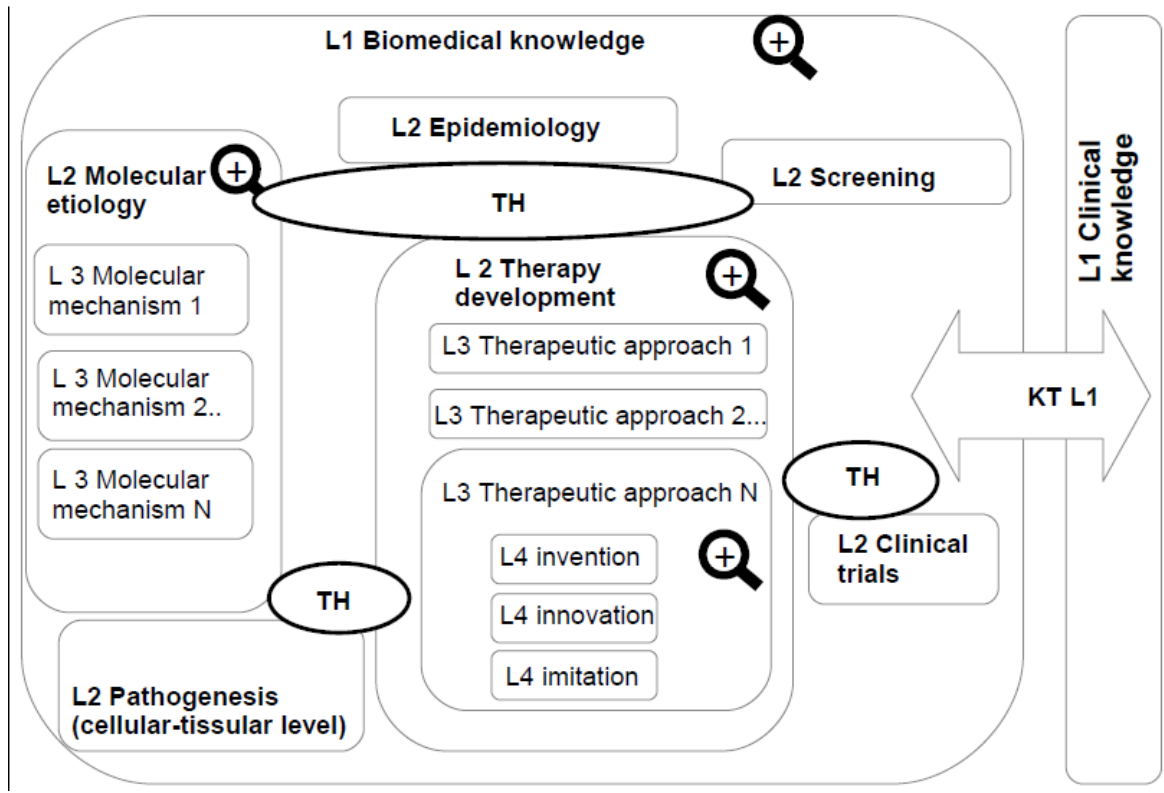

**Figure 2.** Structure of the KT in cervical cancer research and Ebola research.

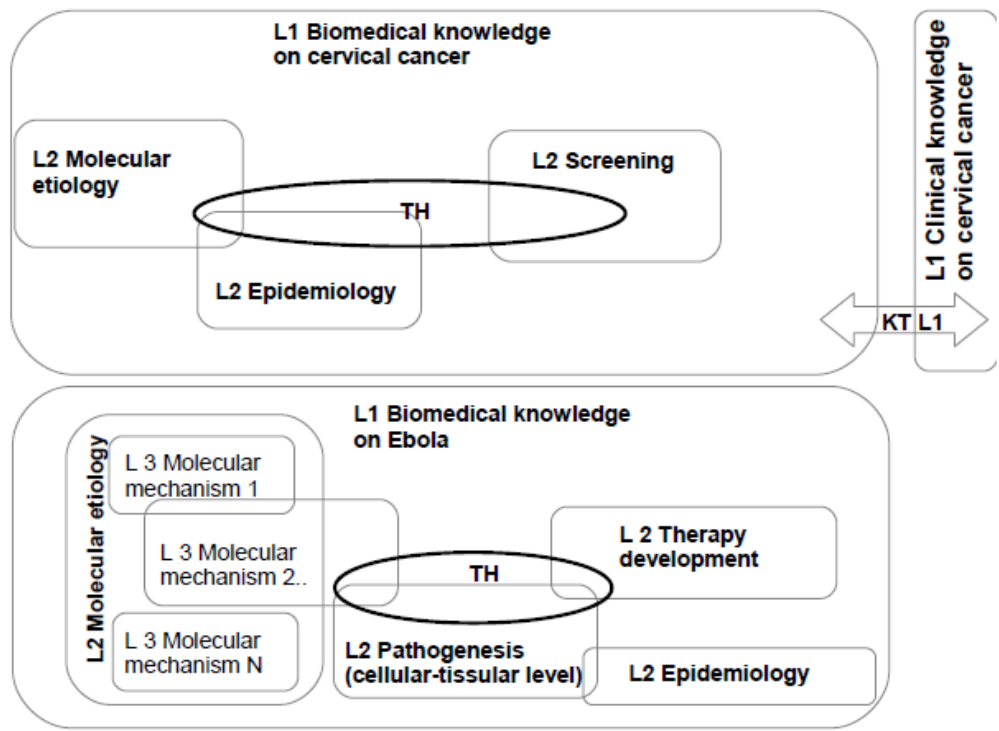